# Using Guided Transfer Learning to Predispose AI Agent to Learn Efficiently from Small RNA-sequencing Datasets

Kevin Li[1,2], Danko Nikolić[3], Vjekoslav Nikolić[3], Davor Andrić[3], Lauren M. Sanders[4,5], Sylvain V. Costes[5]

1. University of California at Berkeley, Berkeley, CA, USA
2. Volunteer Internship Program, NASA Ames Research Center, Moffett Field, CA, USA
3. Robots Go Mental, Frankfurt, Germany
4. Blue Marble Space, Seattle, WA, USA
5. Space Biosciences Division, NASA Ames Research Center, Moffett Field, CA, USA

**Abstract.** Given the increasing availability of RNA-seq data and its complex and heterogeneous nature, there has been growing interest in applying AI/machine learning methodologies to work with such data modalities. However, because omics data is characterized by high dimensionality and low sample size (HDLSS), current attempts at integrating AI in this domain require significant human guidance and expertise to mitigate overfitting. In this work we look at how transfer learning can be improved to learn from small RNA-seq sample sizes without significant human interference. The strategy is to gain general prior knowledge about a particular domain of data (e.g. RNA-seq data) by pre-training on a general task with a large aggregate of data, then fine-tuning to various specific, downstream target tasks in the same domain. Because previous attempts have shown traditional transfer learning failing on HLDSS, we propose to improve performance by using Guided Transfer Learning (GTL). Collaborating with Robots Go Mental, the AI we deploy here not only learns good initial parameters during pre-training, but also learns inductive biases that affect *how* the AI learns downstream tasks. In this approach, we first pre-trained on *recount3* data, a collection of over 400,000 mouse RNA-seq samples sourced from thousands of individual studies. With such a large collection, patterns of expression between the ~30,000 genes in mammalian systems were pre-determined. Such patterns were sufficient for the pre-trained AI agent to efficiently learn new downstream tasks involving RNA-seq datasets with very low sample sizes and performed notably better on few-shot learning tasks compared to the same model without pre-training.

**Introduction**

Given the increasing availability of biological omics data (e.g. transcriptomic, proteomic, etc.) and its high dimensionality and complex nature, there has been growing interest in applying AI/machine learning methodologies to work with such data modalities. Most conventional, state-of-the-art AI algorithms, especially deep learning algorithms, learn best when given a large sample size. However, omics data (such as RNA and DNA sequencing [RNA- and DNA-seq]) is characterized by high dimensionality and low sample size (HDLSS), so current attempts at integrating AI in this domain require significant human guidance and domain expertise in order to mitigate overfitting and generalizability issues. This issue is exacerbated for domains of study focusing on rare diseases or extreme environments such as outer space, which generate very few data points. In such conditions, AI models would ideally learn from small sample sizes without significant human interference, as such interference can induce artificial biases (for example, highly variable gene selection, a conventional method of feature selection in omics studies based solely on variance, tends to omit many predictive genes [1][2][3]).

While it may seem unlikely for deep neural networks to be able to learn from few samples (i.e. "few-shot learning") without severe overfitting and lack of generalizability, one look at biological neural networks shows us that few-shot learning should not be such an unreasonable task for intelligent agents. A typical human toddler only needs a couple examples of cats and dogs to be able to learn a generalizable mental model that allows them to tell the two species apart in a plethora of contexts and configurations. An artificial neural network would require hundreds, if not thousands of samples to achieve the same level of generalizable accuracy. What accounts for this difference?

One explanation is that humans don't have to learn from scratch every time they learn a new, specific task. Instead, we have an agglomeration of general knowledge learned from related experiences that we can utilize to learn new tasks efficiently (i.e. with few training samples). Thus, when the training sample size is small for a particular target task we want an AI agent to learn starting from previous knowledge. Transfer learning is based on such a strategy. In this work, we propose to pre-train an AI agent on a general task to gain sufficient knowledge about biological inter-gene expression relationship by using large aggregate of published omics data. Such an AI agent could then be fine-tuned (starting from the weights learned during pre-training) to answer specific questions about gene expression in smaller, more specialized datasets. Our goal is to significantly increase performance on downstream tasks with small training sizes compared to if one were to start training with randomized weights [4]. However, in the case of space biology omics data, sample size is often less than 100 [5] and traditional transfer learning methods may not be sufficient. In the case of such extreme HLDSS character, it remains to be seen if gaining "general knowledge" alone is sufficient to close the performance gap between AI and human few-shot learning.

To potentially address this pitfall, a new methodology called Guided Transfer Learning (GTL) will be tested and compared to traditional transfer learning. Briefly, AI researchers have proposed another explanation to account for the remaining gap, one that revolves around the idea of meta-learning, or learning *how* to learn. Biological brains have evolved such that they are *predisposed* to learn certain general domains of tasks (e.g. image recognition, language acquisition) very quickly, even *before* they have accumulated significant general knowledge from prior experiences within those domains [6][7]. Finding a way to predispose an AI agent to learn efficiently from certain domains of data without millions of years of evolution would be a promising step toward closing the AI-human gap in few-shot learning. Along this train of thought, a startup called Robots Go Mental developed the new methodology GTL, which allows the AI to not only learn good initial parameters (representing general knowledge of a domain) during pre-training, but also learn inductive biases that affect how the AI learns in the future [5]. Essentially, the AI is learning *how* to learn most efficiently for a particular data domain, not just learning general knowledge about the domain.

GTL has been shown to drastically improve performance on one-shot learning tasks. In the original GTL paper [5], neural networks were pre-trained on the MNIST database (Modified National Institute of Standards and Technology) using both traditional transfer learning and GTL. The resulting models were then trained and tested on the Omniglot dataset (containing examples of characters from over 50 alphabets) using the following scheme to test one-shot learning ability: train with one example of each character, test classification performance on the rest of the dataset. The model pre-trained with GTL had a two to three-fold increase in accuracy compared to the traditional pre-trained model. Both had an accuracy drastically higher than one would expect by chance.

In this project, we collaborated with the creators of GTL (founders of Robots Go Mental) and predisposed a neural network with an scBERT architecture [8] to learn efficiently from mouse RNA-seq data by pre-training using GTL in a self-supervised manner on a large collection of mammalian RNA-seq data sourced from thousands of different studies. The goal was to create an AI agent that is an expert in RNA-seq data, such that it would be able to learn any downstream task involving RNA-seq data even when the training sample size is extremely low. We then compared the few-shot learning performance of the model trained with GTL to those of models trained with only traditional transfer learning and trained from scratch.

**Methods**

To create an AI agent that would be effective at few-shot learning tasks involving low sample size mouse RNA-seq data, we pre-trained a deep neural network architecture using both conventional pre-training and Robot Go Mental's GTL algorithm on a large aggregate of mouse RNA-seq data.

The model architecture we chose was scBERT [8], which is a transformer-based encoder architecture built off BERT [9], a state-of-the-art natural language processing model. BERT-like models are advantageous in that they can be pre-trained on massive amounts of *unlabeled* data in a self-supervised fashion; the model masks certain parts of the input (e.g. expression values of random genes) and is trained to reconstruct the original input from the masked version, the ultimate goal being to learn an informative, usually lower-dimensional encoding of the data that can produce accurate reconstructions back to the original space [9]. In the process, the AI learns more about the internal structure of the data, such as interactions and dependencies between features (e.g. long-range dependencies between words in a sentence in the case of BERT or gene-gene interactions in the case of scBERT). The original scBERT study used the architecture for downstream cell type annotation tasks involving single-cell RNA-seq data [8], but none of the modifications made to the original BERT architecture in scBERT were specific to single-cell data or cell type annotation tasks. Rather, they were intended to make the architecture compatible with gene expression data in general, including bulk RNA-seq data, the modality used for this study. For example, scBERT does not use positional encoding since the ordering of features does not matter for gene expression data, and also uses embeddings for gene identity (gene2vec [10]) and expression values (expression embeddings) designed to be specifically optimized for gene expression information. scBERT also uses token embeddings to represent the continuous gene expression values, which are binned into discrete categories using bag-of-words techniques [11], which has the additional benefit of reducing noise [8]. Additionally, scBERT utilizes a much more scalable approximation of the regular full-rank-attention Transformer called the Performer, which can scale effectively to over 10,000 features/genes (unlike a regular full-rank-attention Transformer which has input length limit of 512) [12].

For the pre-training dataset, we chose *recount3*, a collection of over 400,000 mouse RNA-seq samples sourced from thousands of individual studies [13]. After preprocessing the *recount3* data to prepare it for pre-training (see Data Collection and Preprocessing), the bulk of the project was split into 3 phases, inspired by the approach used in the original GTL paper.

*Phase I: Conventional Pre-training (Traditional Transfer Learning)*

To begin, we pre-trained the scBERT architecture in the conventional, self-supervised way using the entire scBERT encoder-reconstructor architecture on preprocessed *recount3* data (Fig. 1a). The goal was to get the model weights to approach the global optimum for downstream, supervised tasks. The desired result of this phase was a model that had general knowledge about the structure and syntax of gene expression data (e.g. gene-gene interaction) in general, stored in the pre-trained model weights, that would be relevant and transferable for downstream tasks related to RNA-seq data.

*Phase II: Scouting (Guided Transfer Learning)*

We then partitioned the full *recount3* dataset into subproblems for scouting, according to the GTL methodology [5], but k-means clustering was used for partitioning instead of using various category pairs, which could not be applied since our pre-training dataset was unlabeled. After clustering the *recount3* data, we then created subdatasets that contained data from different pairs of clusters. "Scouts" (copies of the pre-trained scBERT architecture) were then trained on these subdatasets to predict the cluster identity of samples. Since these tasks were supervised, the reconstructor portion of the scBERT architecture was replaced with a prediction layer (Fig. 1b). These scout problems were made intentionally easier than the target downstream tasks, since samples in each cluster will automatically be more similar to each other, so that these scouts would converge without too much difficulty. In the process of converging effectively, the scouts learn *guide values* that emphasize *which features* need to be changed a lot in order to achieve effective convergence [5]. Specifically, each scout tracks how much each parameter changes throughout the course of convergence via the squares/absolute difference between the final and initial values of each weight. These differences are then averaged out across scouts and normalized to create guide values; the larger the guide value for a parameter, the more it changed during scout convergence and thus the more it should be allowed to change during fine-tuning. The gradient descent updates during fine-tuning for each parameter are multiplied by these guide values. These guide values then determine how flexible each feature is during downstream fine-tuning to specific tasks (Fig. 2). This is essentially a look-ahead mechanism that scouts "easier versions" of the loss terrain to get a general sense of which features are truly important for global convergence, warning of local optima traps. The goal of this phase was to have the scBERT model learn *how to learn most efficiently* from RNA-seq data during new, downstream tasks, and this knowledge/intuition would be stored in the guide values of the weights in the encoding portion of the scBERT architecture.

*Phase III: Guided Fine-tuning to Supervised Downstream Task*

We then tested the model's ability to learn efficiently for downstream, supervised tasks with small sample sizes. We used the same model architecture as we did in Phase II, except now the model had both pre-trained weights and inductive biases in the form of guide values (Fig. 1c). The model was then trained on NASA OSD-105 [17], a small dataset from the NASA Open Science Data Repository (OSDR) consisting of RNA-seq data taken from the tibialis anterior muscle of 6 spaceflown (FLT) and 6 ground control (GC) mice. The model task was to predict whether a mouse was FLT or GC. We then used two similar datasets, RNA-seq from mouse soleus muscle in OSD-104 [18] and RNA-seq from mouse extensor digitorum longus muscle OSD-99 [19], as validation and test sets, respectively. Each dataset contained mice from a different spaceflight mission; this was intentional so that we could test the ability for the model to generalize to missions that were not represented in the training set (i.e. ensuring the model was not overfitting to any particular mission).

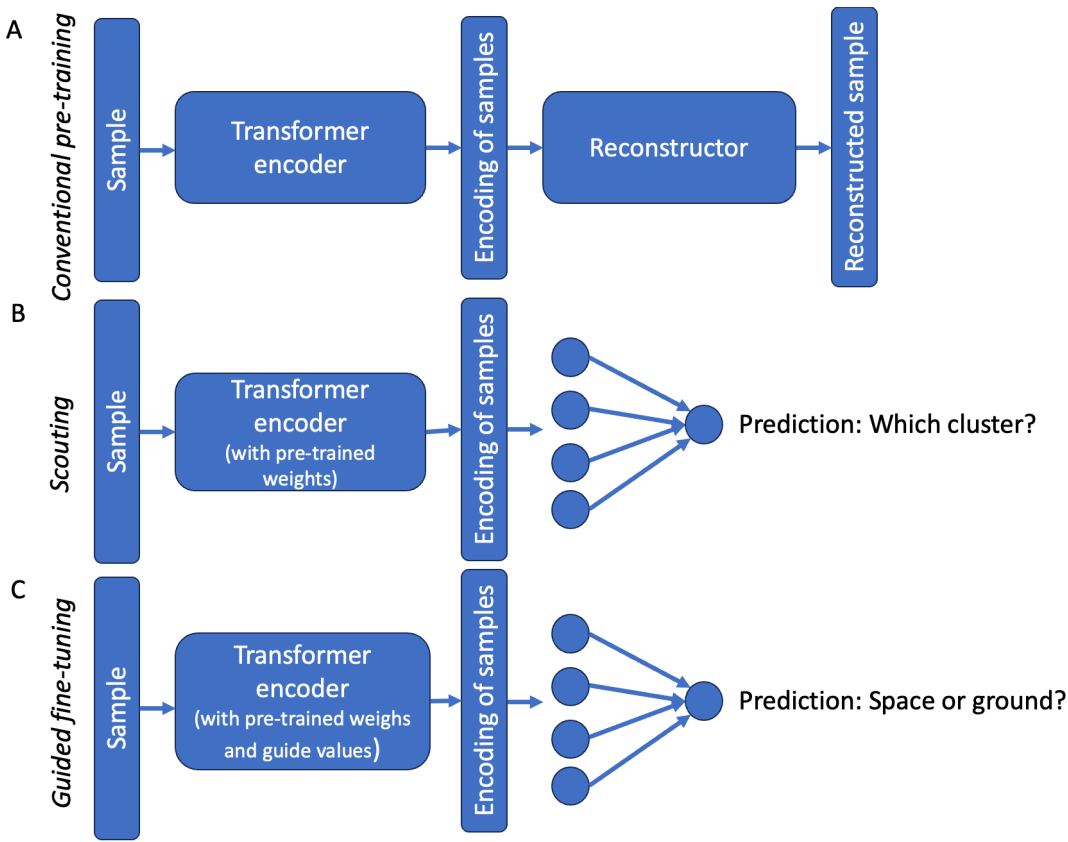

**Figure 1.** High-level overview of scBERT model architectures during various phases of the project. **A)** In Phase 1, the entire encoder-reconstructor architecture of scBERT is pre-trained in a self-supervised manner, and initial weights are learned. **B)** In Phase 2, the entire encoder-reconstructor architecture of scBERT is used during the scouting procedure, and guide values for the weights are learned. **C)** In Phase 3, the encoder portion of the architecture is attached to a fully-connected prediction layer that outputs a label. This architecture is then trained on the downstream supervised task. The reconstructor portion is not used in this phase.

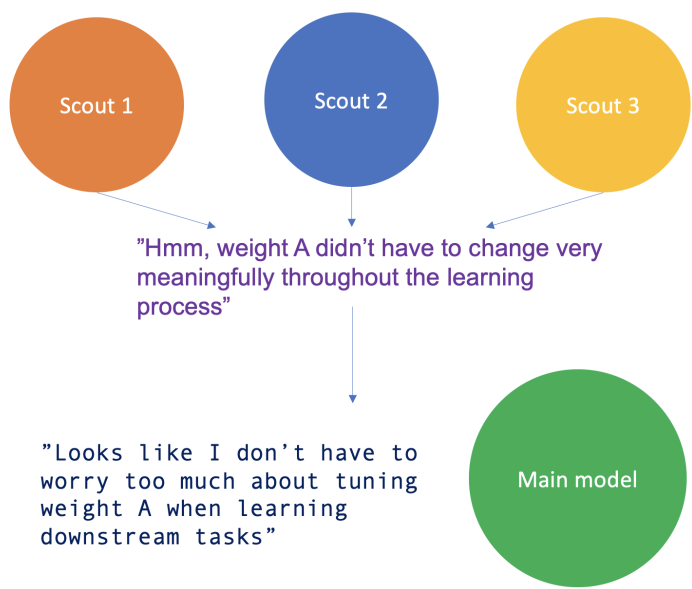

**Figure 2.** Diagram of the scouting procedure. Different copies (or "scouts") of the model architecture are trained on easier subproblems (partitions of the full dataset). During convergence, the scouts keep track of how much each weight in the architecture changes, and this information is averaged across scouts to determine guide values for each weight in the main copy of the model.

**Data Collection and Preprocessing**

*recount3 Preprocessing for Phase I*

The data used in the pre-training and scouting stage (for the purposes of predisposition) consisted of bulk RNA-sequencing data from *recount3* [13], a public database of over 410,235 uniformly preprocessed mouse RNA-seq samples sourced from ~20,000 independent studies, to ensure sufficient number of samples for effective pre-training. Additionally, pre-training with data sourced from multiple datasets will ensure the model is highly generalizable to a variety of downstream tasks/datasets, since it would also be learning to account for technical batch effects, a major problem with biomedical data. If we only used data from a single study, the information learned during pre-training may not be generalizable enough to transfer well to new datasets with different distributions due to technical effects.

The pre-training dataset was preprocessed in order to adjust for distributional artifacts. Many preprocessing steps that are traditionally used in conventional RNA-seq were not employed here due to 1) their infeasibility with enormous amounts of data and/or 2) the lack of generalizability to new data. For example, Trimmed Mean of M-values (TMM) normalization is a robust normalization technique used for single-dataset RNA-seq studies to account for library size differences between samples (as to make expression levels between samples comparable), but it requires calculating size factors for each individual sample based on whole-dataset statistics (i.e. trimmed means) [14][15]. This is expensive and time-consuming to calculate for datasets that are too large to fit in memory (as with *recount3* pre-training data) and must be recalculated each time a new sample is added to the training data, which would force our training dataset to be relatively static to avoid excessive expensive computations. Additionally, the normalization size factors calculated on the training data may not generalize well to downstream datasets. Therefore, we used counts per million (CPM) normalization instead, which accounts for differences in library size by normalizing samples by total expression [15], a slightly less robust but much more feasible method for between-sample normalization given the nature of our problem since it does not require the calculation of whole-dataset statistics. Another common pre-processing step is technical batch effect correction, with methods such as *limma* package's removeBatchEffect function [16], to account for technical differences in the measurement instrument (i.e. the RNA sequencer) between different batches of collected data, making samples comparable between batches. However, these procedures require calculations performed on the entire dataset and must be re-calculated when new data is added to the training data. Additionally, models learned from pre-training data treated with these local batch effect correctors risk not being generalizable to downstream datasets.

Preprocessing the *recount3* data to make it AI-ready involved filtering out samples expressing fewer than 200 genes, CPM normalization, and log2 transformation. These steps mitigated heavy data skew (raw RNA-seq data is sparse and contain many 0 or near-0 values, while log-transformed data distribution is closer to Gaussian) and technical batch/sample effects (i.e. allowed comparison between samples by adjusting for library size differences). We confirmed this by performing PCA with 10 Principal Components (PCs) and visualizing 10% of the data along pairs of these PCs using a pairplot (Fig. 3). Although some batch effects between samples from different studies exist by visual inspection, they did not seem so drastic as to warrant the employment of manual batch effect correction techniques, and scBERT had been shown to generalize well despite batch effects [8].

The PCA plots for the raw data contained many linear artifacts, which indicated that many samples were clustered around the original gene axes (which were rotated during PCA) and suggested the presence of many near-0 values. After preprocessing, however, these linear artifacts were no longer as punctuated and the data seemed to have a much more diffuse, Gaussian-like spread (Fig. 3).

We also color-coded the samples in the PCA plot by whether or not a sample was sourced from a single-cell RNA-seq dataset or a bulk RNA-seq dataset and observed that PC 2 and 3 seemed to separate these two modalities of data pretty significantly, indicating that they may be different enough in nature and should be considered/trained on separately (Fig. 3). Therefore, the scope of the pre-training data was limited to only mouse *bulk* RNA-seq data, since most of the downstream RNA-seq datasets with extremely low sample sizes are bulk RNA-seq data. An additional PCA (with 10 PCs) was performed on mouse *bulk* data alone, color-coded by the original source study the data was extracted from (Fig. 4).

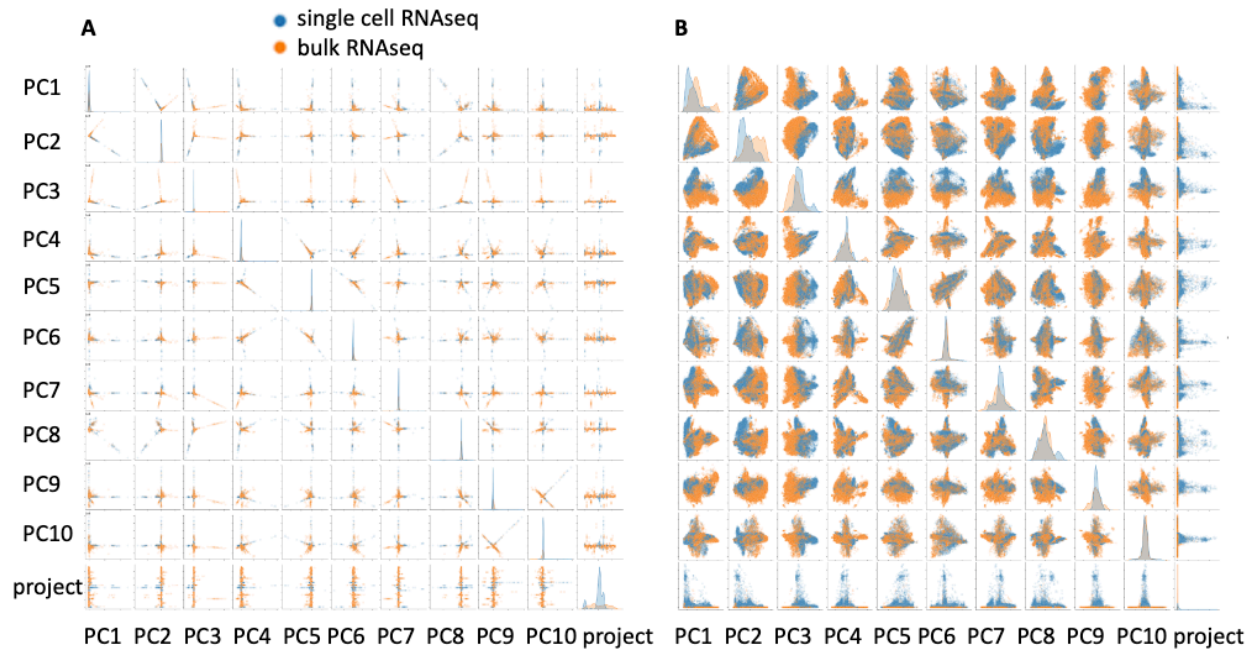

**Figure 3:** PCA pairplots of the first 10 PCs trained on raw (left) and preprocessed (log-transformed, CPM normalized, filtered) *recount3* mouse data (right). Samples are color-coded by their RNA-seq modality (blue for single-cell, orange for bulk).

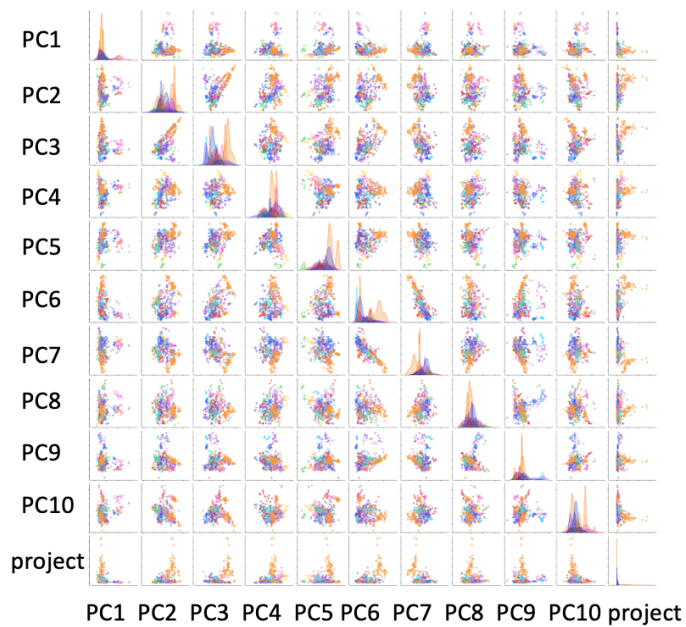

**Figure 4:** PCA pairplots of the first 10 PCs trained on preprocessed *recount3* mouse *bulk RNA-seq* data. Samples are color-coded by their source study.

Only 15,117 genes out of over 30,000 were kept as features from the *recount3* data. These features had corresponding embeddings which were included in the final dataset. We coined such gene features embeddings *gene2vec* and they were extracted from a deep learning ~~morning~~ method similar to *word2vec* in order to encode generalizable information about gene-gene interactions to make scBERT pre-training more efficient [8][10]. At this point, the *recount3* data was ready to be used for scBERT pre-training.

*recount3 Preprocessing for Phase II*

In this phase, we trained scouts on an easy a verifiable subproblems: i.e. the accurate classification of a gene expression profile to a given cluster. This was done by first creating these clusters out of the *recount3* bulk RNA-seq data using K-means in PC space (top 40 PCs were used). Subproblems were created by selecting random pairs of clusters and training the scBERT model to predict the cluster identity of samples as the labels with an 80-20 train-test split (Fig. 6). For these subproblems to be easily achievable, we had to determine the clustering granularity, where each partition had enough data while clusters were clearly distinguishable in the original feature space, so that the scout would be able to converge towards true ground truth. Briefly, having a subproblem with too little training data would be counterproductive toward the goal of creating tasks that can easily be converged on. We plotted our clustering results for K = 5, 10, 15, 20, 30, and 40 using UMAP dimensionality reduction [20] (Fig. 5). We chose maximum K = 40 to ensure all clusters would have an average size greater than 5,000 samples. Based on the plots, we chose K = 30, since at this granularity, most visually distinct groups were classified as separate clusters and individual visual groups were mostly not split across multiple clusters. This also meant each cluster would contain around 6,500 samples on average, meaning each pair of clusters (i.e. training dataset for each scout), would contain around 13,000 samples on average, a reasonable sample size.

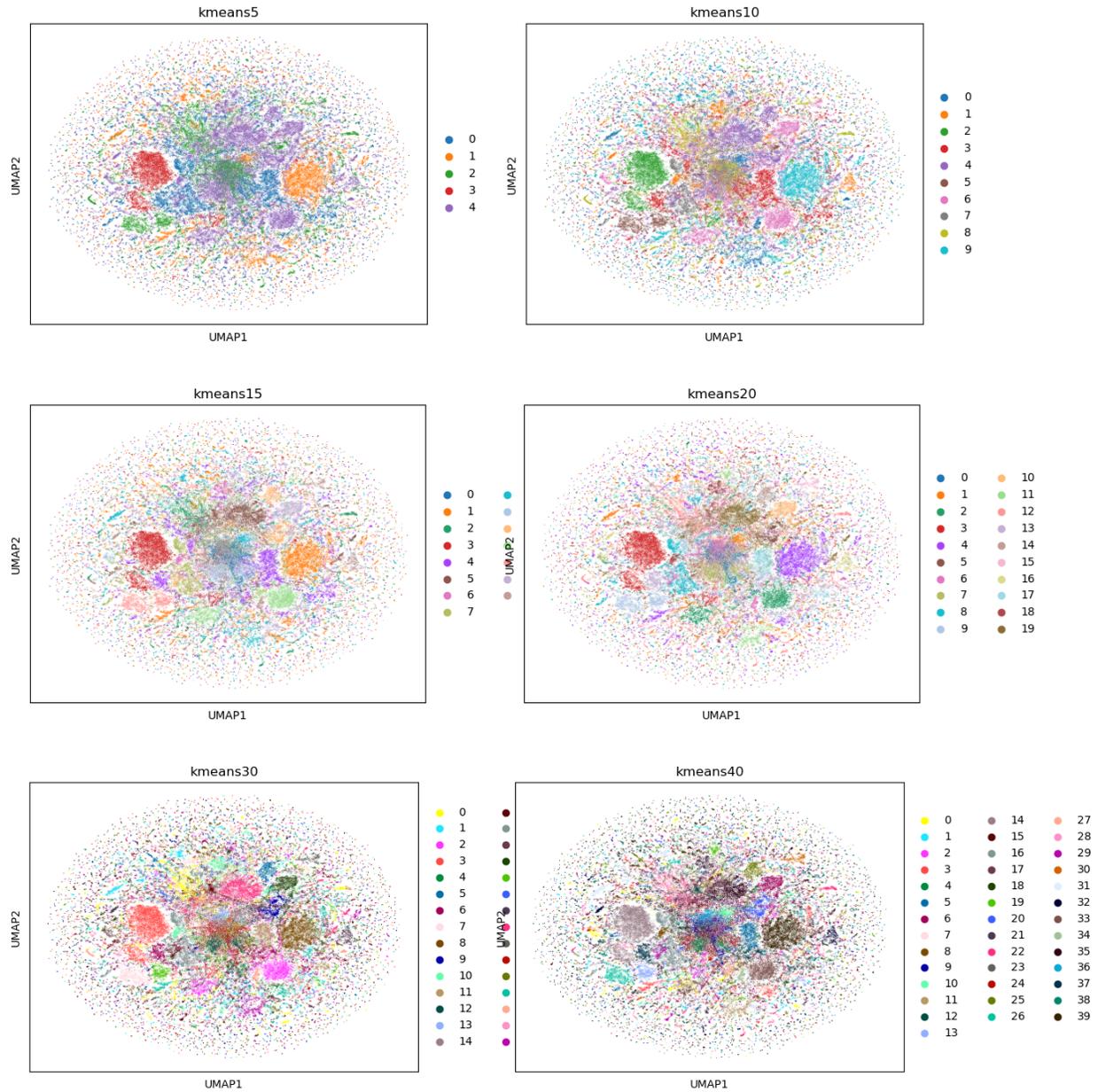

**Figure 5.** *recount3* data was clustered via K-means clustering in PC space (top 40 PCs) and visualized using UMAP. These plots show the results for K = 5, 10, 15, 20, 30, and 40.

Cluster *recount3* data

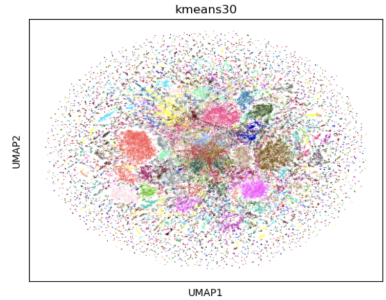

Split full *recount3* data
into pairs of clusters

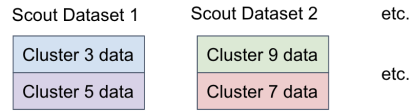

Train scouts to predict
cluster identity of samples
(binary classification)

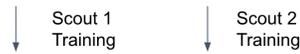

**Figure 6.** After *recount3* data was clustered, subdatasets were created using data from pairs of clusters. Scouts were then trained on these subdatasets to predict the cluster identity of samples (binary classification).

*Downstream Dataset Preprocessing for Phase III*

The datasets used to test the pre-trained model's few-shot learning capacity on supervised, downstream tasks were NASA OSD-105 (mouse tibialis anterior RNA-seq data) [17], OSD-104 (mouse soleus RNA-seq data) [18], and OSD-99 (mouse extensor digitorum longus RNA-seq data) [19]. We fed these datasets through the same preprocessing pipeline (low expression filtering, CPM normalization, and log2 transformation) as the pre-training data so that knowledge from the pre-trained model could be transferable to this downstream supervised task. The downstream task was to differentiate between FLT and GC mouse RNA-seq samples.

**Results**

To evaluate the improvement offered by conventional pre-training and scouting methodologies for few-shot learning tasks, we trained three separate models: 1) a model that did not undergo any pre-training at all (randomly initiated parameters), 2) a model that only underwent conventional pre-training on the full dataset (Phase I), and 3) a model that underwent the full GTL pipeline, with both conventional pre-training on the full dataset and scouting on sub-clusters of the dataset (Phases I and II). The few-shot learning capability of these models were all evaluated with the previously described task for Phase III; we fine-tuned each model on OSD-105 data to differentiate between GC and FLT mice, validated using OSD-104 data during hyperparameter optimization, then tested the predictive accuracy on OSD-99 data. A cosine annealed warm restart learning rate scheduler was employed [21].

According to the loss and accuracy curves for the training and validation data during downstream fine-tuning, we observed that the model with no pre-training at all started overfitting the training set almost immediately, as indicated by the validation loss beginning to increase from the very beginning while the training loss kept decreasing (Fig. 7a). With the conventionally pre-trained model, on the other hand, we observed that before overfitting started occurring, there was a sweet spot at around 200 epochs where the validation loss was at a low and the validation accuracy was at a high of 91% (Fig. 7b), demonstrating that pre-training provided a regularizing effect on the model that mitigated the severe overfitting that occurred in the model with no pre-training. We saw a similar effect with the model that underwent both conventional pre-training and scouting, but the maximum validation accuracy was higher at 100%, and this maximum validation accuracy was maintained for more epochs before overfitting started occurring (Fig. 7c). However, the validation performance was also noisier throughout fine-tuning compared to the other two models.

On the test set, the model with no pre-training performed abysmally, as expected, with a test accuracy of 50%. Both pretrained models achieved a relatively high test accuracy of 83% (Table 1).

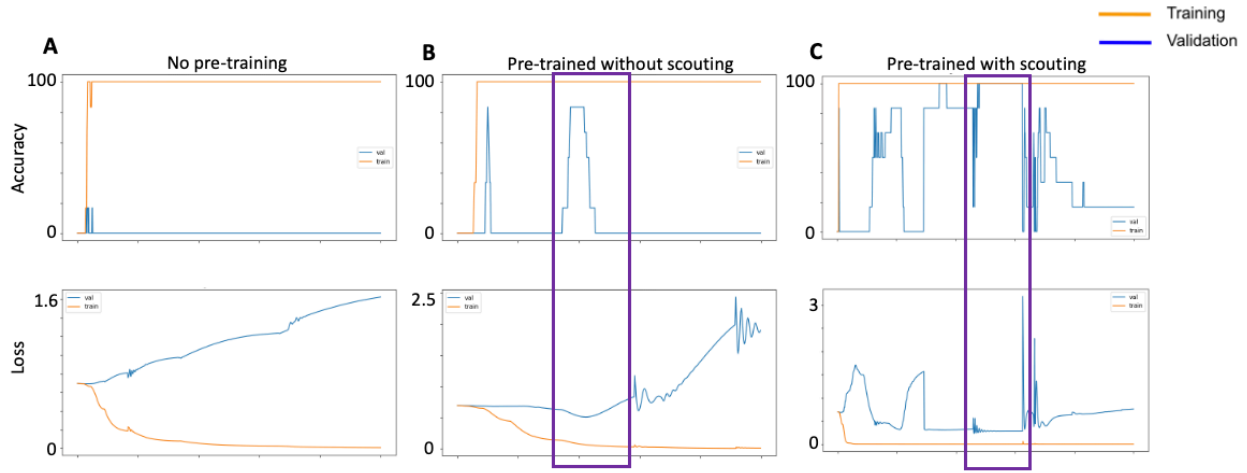

**Figure 7.** Training and validation accuracy (top) and loss curves (bottom) for: **a)** a model with no pre-training, **b)** a model that was only conventional pre-trained without scouting, and **c)** a model that was conventionally pre-trained with scouting (full GTL pipeline). The model with no pre-training started overfitting immediately, while the pre-trained models both achieved strong validation performance before overfitting occurred (purple boxes).

| Experiment | OSD-99 Test Accuracy |
|---|---|
| 1. No pre-training | 50% |
| 2. Conventional pre-training, no scouting | 83% |
| 3. Conventional pre-training with scouting (GTL) | 83% |

**Table 1.** Few-shot learning test accuracies of 1) a model with no pre-training, 2) a model that was only conventional pre-trained without scouting, and 3) a model that was conventionally pre-trained with scouting (full GTL pipeline). The models were trained on a 12-sample NASA OSDR dataset (OSD-105) and tested on another 12-sample NASA OSDR dataset (OSD-99). The pre-trained models performed notably better on the test set than the model without any pre-training.

**Discussion**

The results of Phase III fine-tuning show that scBERT pre-trained on *recount3*, in one form or another, provides significant improvement in few-shot learning performance compared to scBERT without any pre-training. The advantage of pre-training could perhaps be best described as a regularizing effect, as indicated by a sweet spot in the validation loss and accuracy curves, where the model is both accurate and generalizable, that occurs before the model starts overfitting. This improvement in generalizable accuracy provided by pre-training is also reflected in the test set results.

In terms of the benefit that GTL provides over conventional transfer learning, the validation performance suggests that scouting has the potential to provide improvement. The maximum achieved validation accuracy was not only higher with the scouted model, but this maximum validation performance was maintained throughout over three times as many epochs compared to the model that only underwent conventionally pre-training, suggesting that the maximal performance achieved by the scouted model is more robust and less likely due to random noise. However, we do see that the validation curves for the scouted model were slightly noisier overall throughout the course of fine-tuning. This is likely a result of the cosine annealed warm restarts learning rate scheduler, which cycles through learning rates to "jump out of" local minima [21], so the jumps in validation accuracy and loss in the beginning are likely points where the model escapes a local minimum.

The test results, on the other hand, suggests that additional refinement is needed for identifying and selecting scouts. For example, future work could focus on creating scouts which share biological information such as gene networks or transcription factor control. It is important to keep in mind that the test set used is quite small (12 samples), and a larger, more robust test set may better demonstrate the differences in performance between models trained with and without scouting. Additionally, a more difficult downstream task could have also better differentiated the two pre-trained models. The model that only underwent conventionally pre-trained already performed very well on the validation and test sets, so it is reasonable that GTL perhaps did not show drastic improvement because there was just not much room for improvement. Potential, harder tasks could include testing the generalizability of the model to different types of tissues, as OSD-105, OSD-104, and OSD-99 all contain data for muscle-related tissues only.

**Conclusion**

Given the relatively strong generalizability of scBERT models pre-trained with bulk RNA-seq data from *recount3*, we believe these models are potential candidates for foundation models that can be applied to various downstream, RNA-seq-related tasks cursed with small training sets. Such foundation models will drastically change the landscape of space omics research, an area currently characterized by extremely small datasets, excessive human tinkering of data, and, consequently, severe problems with the generalizability of insights gained from analysis. Having an AI model that is predisposed with general knowledge about bulk RNA-seq data and correct inductive biases for how to learn most efficiently in this domain would mitigate the notorious challenges with working in this field. Future work could train similar models for other omics data types to learn foundational, complex biological relationships and then be applied to specific problems in smaller datasets. In general, the GTL approach shows promise for complex biological datasets and future efforts will expand the flexibility and applicability of this approach.

**Ethical Statement**

Any project involving artificial intelligence and genomic data has heavy ethical implications concerning personal identification, privacy, and providing potential bases for discriminatory mindsets. This project itself only involves animal data, so human identification and privacy is not of direct concern during the research phase; however, one of the ultimate goals of space biology research is to apply the developed technologies on human astronauts. During the translation process from research to human healthcare, it is important to keep in mind the potential intrusions of privacy and exacerbation of discrimination that may come with using subjects' biological data and to properly mitigate and account for those risks.

**Code Availability and Documentation**

Code is available at this Github repository: https://github.com/kevinli5941/RNALearner

The repo consists of my own code, as well as code from scBERT's GitHub:

https://github.com/TencentAILabHealthcare/scBERT

Reproducibility workflow:

https://docs.google.com/document/d/1IKta_aGRnhv4w_rLRJq8dGJsn1SNCpQGu4m-8GHwDhI/edit?usp=sharing